# Optimal Operation of a Tidal Lagoon as a Flexible Source of Electricity

Tong Zhang, Christopher Williams, Reza Ahmadian and Meysam Qadrdan*
School of Engineering, Cardiff University, Cardiff, United Kingdoms
Email: ZhangT44@cardiff.ac.uk, ChristopherJohnWilliams1@gmail.com, ahmadianr@cardiff.ac.uk, Qadrdanm@cardiff.ac.uk

*Abstract*—As the demand for electricity and the need for power systems flexibility grow, it is crucial to exploit more reliable and clean sources of energy to produce electricity when needed most. Tidal lagoons generate renewable electricity by creating an artificial head difference between water levels on the seaside, driven by tides, and water levels inside the basin, controlled by flow through the structure. Depending on the level of seawater, power generation from a tidal lagoon can be controlled, i.e. shifting power generation in time. This paper aims to investigate the operation of a tidal lagoon in response to fluctuating electricity prices. By developing an optimal operation model of a tidal lagoon, its schedule in the day-ahead wholesale electricity market was optimized to achieve maximum revenue. The Swansea Bay tidal lagoon was used as a case study. It was demonstrated that by exploiting the flexibility offered by the tidal lagoon, it can achieve a higher revenue in the day-ahead market, although their total electricity generation is reduced.

## I. Introduction

The United Kingdom faces an unprecedented challenge to provide reliable and clean energy in the coming decades. Many renewable energy sources (RESs), such as wind and solar energy, have been deployed to satisfy power demands with less greenhouse gas emissions. However, the highly variable and unpredictable nature of these RESs often brings difficulties to the economical and secure operation of the power system. In contrast, tidal energy is known for its high predictability. Many tidal-related generation technologies can provide energy with less uncertainty, which could act as a reliable energy source for the power system. As the demands for power supply and system flexibility increase, it is wise to bring more tidal energy into the energy industry.

Tidal lagoon, a less-exploited power generation technology, has been brought up several years ago as an alternative to tidal barrages. It captures the potential energy from the water level difference and generates electricity. Although the piloting project (the Swansea Bay project) was rejected by the British government in 2017, a new project at the same location has just been announced in October 2021. This implies great confidence from the industry in this generation technology.

Many studies on the design, control, simulation and analysis of tidal lagoons and their precedents have been carried out. Studies on tidal energy resources, especially around the UK's western coastal area have been carried out decades ago [1], as well as the environmental and social impacts of this technology [2]. Designs of tidal barrages, which are also used by tidal lagoons, have been developed in the 1980s [3]. There are also numerical models (e.g., 0-D model and 2-D model) that have been developed to simulate the optimal operation of tidal lagoon [4-6]. The contribution of tidal energy to the future of the UK energy supply mix has been assessed, too [7]. The existing research focuses on the tidal lagoon itself and normally pursues the maximisation of generation output. There hasn't been much discussion on how the tidal lagoon will integrate into the power system and react to the energy market. There are still some research questions that need to be investigated further: How much energy should a tidal lagoon provide during each time period? How to ensure a tidal lagoon generates power when energy is most needed, instead of simply maximising energy production? How much energy can a tidal lagoon store and how does the energy storage change with the sea level going up and down?

To address some of the aforementioned issues and lay the foundation for future studies on tidal lagoons, this paper explores the optimal operation scheme of a tidal lagoon in response to wholesale electricity prices. By constructing an optimisation problem with the goal of the maximum revenue from power generation, a day-ahead operation scheme of a tidal lagoon is obtained. The head differences and number of active turbines are adjusted by the tidal lagoon during the operation to reach the optimum operating schedule. The energy stored by a tidal lagoon is also quantified in this paper to further reveal the contribution of a tidal lagoon to the power system.

The remaining sections of this paper are organized as follows: Section II will give a brief introduction to the structure and operation scheme of the tidal lagoon; Section III will present the optimisation model for the tidal lagoon; Section IV will present the results and data analysis on a test run using the Swansea Bay Tidal Lagoon case; Section V will conclude this paper and provide directions for future work.

## II. Introduction to Tidal Lagoon

A tidal lagoon is an artificial reservoir built in the coastal area to generating an artificial head difference between the open sea and impounded area. Turbines and sluice gates embedded within the structure walls are mainly used to generate electricity and control the water level inside the lagoon, respectively. The aerial and cross-sectional views of a proposed tidal lagoon, namely the Swansea Bay tidal lagoon, are illustrated in Fig. 1 and 2. As the water rushes through the turbines due to the height difference between the water on each side of the lagoon, the turbines are forced to rotate and then convert the kinetic energy into power. The sluices are openings in the lagoon's wall, which can be opened and closed and are designed to allow large volumes of water to pass into or out of the lagoon in a short period of time.

As the tidal lagoon generation depends on the rise and fall of the tides, the lagoon can be operated using two common operation schemes as follows: one is '*one-way generation*' (generating energy with ebbing tides or flooding tides only)

The authors would like to thank EPSRC for supporting this research through funding MISSION project (EP/S001492/1).

XXX-X-XXXX-XXXX-X/XX/$XX.00 ©20XX IEEE

and the other is '*two-way generation*' (generating energy with ebbing and flooding tides both). In this paper, the ebb-generation (a form of one-way generation) is investigated. During flood tide, the lagoon is filled through the sluice gates and turbines to achieve maximum water level inside the lagoon, which is named as the filling phase. Followed by the holding phase, when the turbines and gates are closed at highwater to create a head difference across the lagoon. Then the generating phase starts, as the turbines open and the water is released through the turbines when a predetermined head difference is reached. As the turbines rotate due to the rush of water, they convert kinetic energy into electricity. When the water level inside the lagoon is no longer high enough to support the turbine generation, the turbines will stop generating. This process can be seen in Fig. 3.

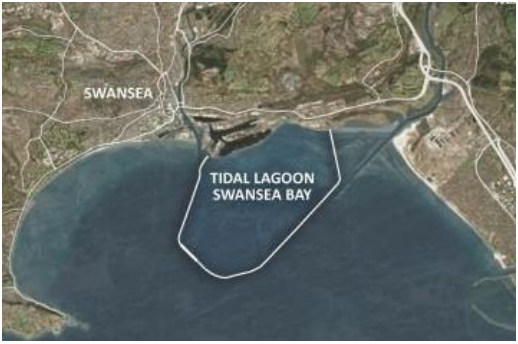

Fig. 1. The aerial view of future Swansea Bay Lagoon [8]

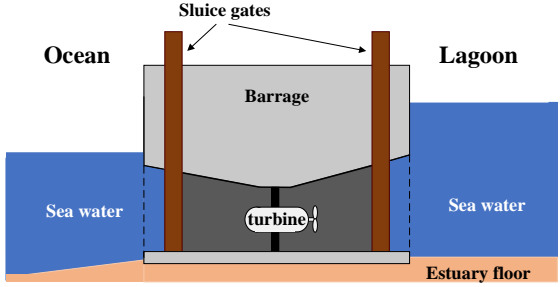

Fig. 2. The structure of a tidal lagoon's wall

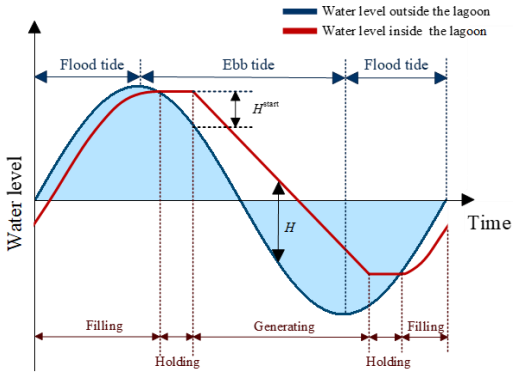

Fig. 3. The ebb-generation scheme of a tidal lagoon

### III. THE FORMULATION OF TIDAL LAGOON'S OPTIMAL OPERATION SCHEME

#### A. Formulation of optimisation problems

In this paper, we formulated the operation of a tidal lagoon as two different optimisation problems. In the first optimisation problem, the objective function is to maximise the generation of electricity over 24 hours, as shown by (1). The objective function of the second optimisation problem is to maximise the revenue of the tidal lagoon, over a similar period, by selling its electricity to the day-ahead wholesale electricity market, as shown by (2).

$$\max \sum_{t \in \mathcal{T}} \left( E_t^{\text{TL}} \right), \quad (1)$$

In (1), $E_t^{\text{TL}}$ represents the energy production (MWh) of the tidal lagoon at time $t$; $\mathcal{T}$ is the index collection of time periods.

$$\max \sum_{t \in \mathcal{T}} \left( C_t^{\text{E}} \cdot E_t^{\text{TL}} \right), \quad (2)$$

In (2), $C_t^{\text{E}}$ represents the electricity price (£/MWh) at time $t$.

The constraints at time $t$ are formulated based on a 0-D representation of a tidal lagoon [9] as below:

$$z_t^{\text{in}} = \begin{cases} z_t^{\text{out}} & t=0 \\ z_{t-1}^{\text{in}} - \dfrac{Q_t^{\text{TL}}}{A}\Delta T & t \geq 1 \end{cases}, \quad (3)$$

$$H_t = z_t^{\text{in}} - z_t^{\text{out}}, \quad (4)$$

$$Q_t^{\text{TL}} = \delta_t^{\text{F}} \left( n^{\text{S}} Q_t^{\text{S}} + \sum_{i \in S^{\text{T}}} Q_{i,t}^{\text{T-FILL}} \right) + \sum_{i \in S^{\text{T}}} \delta_{i,t}^{\text{G}} \cdot Q_{i,t}^{\text{T-GEN}}, \quad (5)$$

$$P_t^{\text{TL}} = \sum_{i \in S^{\text{T}}} \delta_{i,t}^{\text{T}} \cdot P_{i,t}^{\text{T}}, \quad (6)$$

$$E_t^{\text{TL}} = P_t^{\text{TL}} \cdot \Delta T, \quad (7)$$

where $z_t^{\text{in}}$ and $z_t^{\text{out}}$ represent the water head levels (m) inside and outside the tidal lagoon, respectively; $Q_t^{\text{TL}}$ represents the total flow rate of water (m³/s) flowing through the tidal lagoon; $\Delta T$ is the time step (h); $A$ is the surface area of the tidal lagoon (m²); $H_t$ represents the head difference between the inside and outside (m), which stays between -2 m and 8 m for the case studied in this paper; $Q_t^{\text{S}}$ represents the flow rate of water (m³/s) flowing through sluice gates; $Q_{i,t}^{\text{T-FILL}}$ and $Q_{i,t}^{\text{T-GEN}}$ represent the flow rate of water (m³/s) flowing through each turbine during the filling phase and the generating phase, respectively; $n^{\text{S}}$ is the number of sluice gates; $\delta_t^{\text{F}}$ is a binary variable indicating whether the tidal lagoon is at filling phase (during the filling phase, $\delta_t^{\text{F}} = 1$; otherwise, $\delta_t^{\text{F}} = 0$); $\delta_{i,t}^{\text{G}}$ is a binary variable indicating the status of the $i$ th turbine (when the turbine is operating and generating electricity, $\delta_{i,t}^{G}=1$; otherwise, $\delta_{i,t}^{G}=0$); $S^T$ is the index collection of turbines; $P_{i,t}^{T}$ represents the power output (MW) of the $i$ th turbine; $P_{t}^{TL}$ and $E_{t}^{TL}$ represent the power output (MW) and energy production (MWh) of the whole tidal lagoon, respectively. When the water is flowing out of the tidal lagoon, $Q_{t}^{S}$, $Q_{i,t}^{T-FILL}$, $Q_{i,t}^{T-GEN}$ and $Q_{t}^{TL}$ are positive; when water enters the lagoon, these variables stay negative.

Constraint (3) sets the water level inside the tidal lagoon at the initial time period and relates $z_{t}^{in}$ of other time periods to the amount of water flowing out of the tidal lagoon. Constraint (4) defines the head difference in the tidal lagoon operation. Constraint (5) defines the total amount of water flowing through the tidal lagoon. Constraints (6) and (7) pertain to the energy generation, defining the power output and energy production of the whole tidal lagoon. The calculation for $Q_{t}^{S}$, $Q_{i,t}^{T-FILL}$, $Q_{i,t}^{T-GEN}$ and $P_{i,t}^{T}$ are given as follows.

### B. Linearisation of the optimisation problem

The water flow through sluice gates and turbines during the filling phase could be calculated by nonlinear orifice equations:

$$Q_{t}^{S} = C^{S} A^{S} \sqrt{2gH_{t}}, \quad (8)$$

$$Q_{i,t}^{T-FILL} = C^{T} A^{T} \sqrt{2gH_{t}}, \quad (9)$$

where $C^S$ and $C^T$ are the discharge coefficients, normally set as 1.0; $A^S$ and $A^S$ are the cross-sectional flow area of a sluice gate/turbine; $g$ is the gravitational acceleration (m/s$^2$). For the tidal lagoons with maximum head difference lower than 7 m, Equations (8) and (9) could be approximated as linear equations in order to reduce computation complexity:

$$Q_{t}^{S} = K^{S} A^{S} H_{t}. \quad (10)$$

$$Q_{i,t}^{T-FILL} = K^{T} A^{T} H_{t}. \quad (11)$$

where $K^S$ and $K^T$ are the fitted coefficients used to approximate the nonlinear equations.

Hill charts have been used to estimate power generation of and flow through turbines during the generating phase corresponding to different head differences across the turbine [1,4], as shown in Fig. 4. Both $Q_{i,t}^{T-GEN}$ and $P_{i,t}^{T}$ start increasing when $H \geq H^{min}$. $Q_{i,t}^{T-GEN}$ reaches the maximum when $H_t = 7$ m and decreases as the head difference increases, while $P_{i,t}^{T}$ steadily climbs until it meets the rated power capacity.

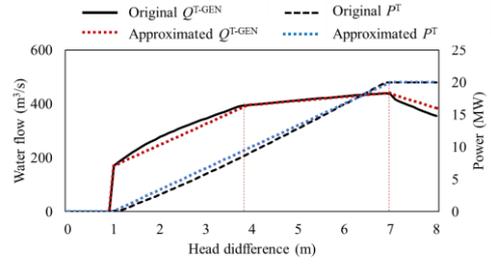

Fig. 4. The hill cahrt for turbine flow rate and power generation

Based on the curves of this chart, $Q_{i,t}^{T-GEN}$ and $P_{i,t}^{T}$ were also approximated:

$$Q_{i,t}^{T-GEN} = \begin{cases} 0 & 0 \leq H_t < H^{min} \\ 92.99+77.60H_t & H^{min} \leq H_t < 3.9 \\ 337.60+14.81H_t & 3.9 \leq H_t < 7 \\ 807.19-52.83H_t & 7 \leq H_t \leq 8 \end{cases} \quad (12)$$

$$P_{i,t}^{T} = \begin{cases} 0 & 0 \leq H_t < H^{min} \\ -3.33+3.33 H_t & H^{min} \leq H_t < 7 \\ 20 & 7 \leq H_t \leq 8 \end{cases} \quad (13)$$

To transform the optimisation problem into a MILP problem, Constraints (12) and (13) were built in a linear form using binary variables. For (5) and (6), the bilinear terms were linearised as well. Taking (5) as an example, the replacement constraints of $Q_{i,t}^{T-GEN}$ are written as below:

$$Q_{t}^{TL} = \delta_{t}^{F}\left(n^{S}Q_{t}^{S} + \sum_{i \in S^T} Q_{i,t}^{T-FILL}\right) + \sum_{i \in S^T} z_{i,t}^{T-GEN}, \quad (14)$$

$$z_{i,t}^{T-GEN} \leq \delta_{i,t}^{G} \cdot \overline{Q}_{i,t}^{T-GEN}, \quad (15)$$

$$z_{i,t}^{T-GEN} \leq Q_{i,t}^{T-GEN}, \quad (16)$$

$$z_{i,t}^{T-GEN} \geq Q_{i,t}^{T-GEN} - \left(1-\delta_{i,t}^{G}\right) \cdot \overline{Q}_{i,t}^{T-GEN}, \quad (17)$$

$$z_{i,t}^{T-GEN} \geq 0, \quad (18)$$

where $z_{i,t}^{T-GEN}$ is a new ancillary variable introduced to linearise the term $\delta_{i,t}^{G} \cdot Q_{i,t}^{T-GEN}$, and $\overline{Q}_{i,t}^{T-GEN}$ is the maximum bound of $Q_{i,t}^{T-GEN}$. In this way, the optimisation model was fully converted into a mixed integer linear problem, which saves a considerable amount of computation time.

## IV. CASE STUDY

To show the result generated by the proposed optimisation model of operation scheme, a test run was carried out using the Swansea Bay Lagoon as a case study. This test case is a tidal lagoon with rated capacity of 320 MW (containing 16 turbines with maximum generation capacity of 20 MW each and $H^{min}$ set to 1 meter). The surface area of the tidal lagoon is 11.5 km² while the total area of sluice gates is 800 m². The optimisation problem aims to provide a 24-hour operation scheme with a 30-min time step, which was modelled in Python and solved by Gurobi. Detailed results and data analysis are discussed in the following sections.

*A. The result of maximum electricity generation*

To demonstrate and analyse the proposed optimisation model, the results of the one-day operation scheme for maximum electricity generation are demonstrated in Fig. 5 to 8. Fig. 5 shows the water levels and head differences during the operation. When the ebb tide begins, both turbines and sluices are closed until the head difference is larger than $H^{min}$. Then the lagoon starts the generating phase, during which a certain number of turbines are generating and releasing seawater to the sea. In this case, the lagoon starts the generation several hours after the beginning of the generating phase, so the turbines could make use of higher head difference and achieve the maximal energy generation during the one-day optimisation. The flow rates through the turbines and sluices are presented in Fig. 6. When turbines are generating electricity, the sluices are closed and the water runs through the turbines only; during the flood cycles, both turbines and sluices are open to ensure the lagoon can be quickly filled to the highest level and ready for next generation phase. The power generation, revenue and price profile are demonstrated in Fig. 7. As the power production is determined by head difference, these two shares the same trend throughout the operation. The number of active turbines is demonstrated in Fig. 8. At the beginning of a generating phase, the head difference is small and there are fewer operating turbines. As $H_t$ increases, more turbines are activated to maximise the power output of the tidal lagoon.

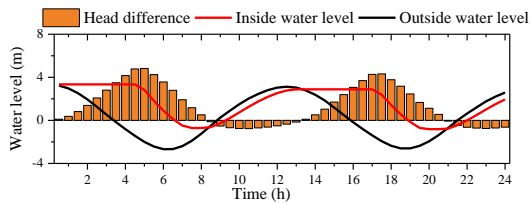

Fig. 5. Head difference and water levels

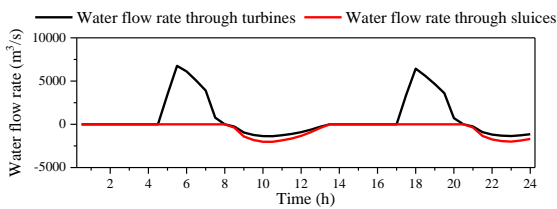

Fig. 6. Total amout of water flowing thourough turbines and sluices

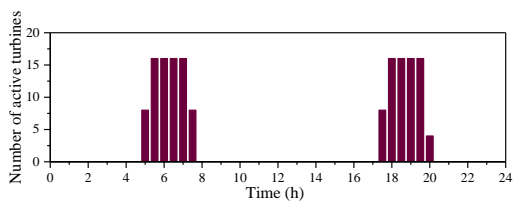

Fig. 7. Number of active turbines

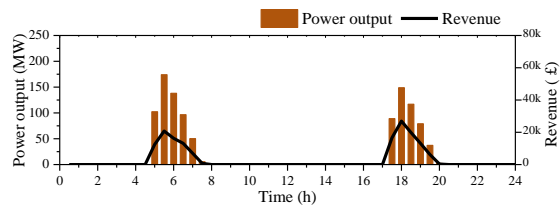

Fig. 8. Power output and revenue

*B. Comparison between different electricity price profiles*

To reach maximum revenue, the power generation of a tidal lagoon can be adjusted by controlling the number of active turbines. To show how tidal lagoon reacts to the fluctuation of energy prices, the optimal schemes of the same tidal lagoon (with the same tidal pattern) under different price profiles [10] were calculated and analysed. The result from maximum electricity generation in Section A was used as the benchmark to demonstrate tidal lagoon's response to electricity price.

The results of the proposed optimisation model under two different price profiles are demonstrated in Fig. 9 and Fig. 10. When in pursuit of maximum revenue, the tidal lagoon adjusts the number of active turbines to control the amount of electricity generated, which subsequently influences the amount of water released back to the sea and the speed of lowering the water level inside the lagoon.

The optimal scheme in Fig. 9 was generated under the same price profile as the benchmark in Section A. By adjusting the on/off status of turbines, the tidal lagoon could increase its energy production when electricity price is high. During the first generating phase, the number of active turbines gradually increases to the maximum at Hour 6, instead of staying the maximum during most of the time (compared with Fig. 7). This is because the electricity price is higher after the Hour 6, a subset of turbines are shut and started until the price climbs. Similarly, as the electricity price drops during the second generating phase, the generation time window has been moved an hour earlier so that more electricity could be generated at higher price and the tidal lagoon could gain more revenue. The comparison of total energy production and revenue is given in Table I. Though the operation scheme for maximum revenue generates less energy than the benchmark, it still gains more revenue by shifting the electricity generation to the time periods with higher electricity price.

Fig. 10 demonstrates the operation scheme in response to another price profile. The turbines start earlier to catch the higher electricity price before Hour 5. In the second generation phase, the electricity price rises to the highest around Hour 18 to Hour 19. Compared with Fig. 7 and Fig. 9, the turbines are switched on later until the electricity price is greater, when a much higher power output could be seen at Hour 18. In this way, the tidal lagoon could gain a higher revenue than the maximal energy generation scheme. This also shows the flexibility of the tidal lagoon to move the generation time window or adjust the generation time duration in response to the price changes.

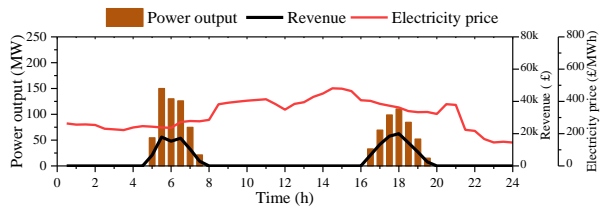

(a) Power output, revenue and price profile

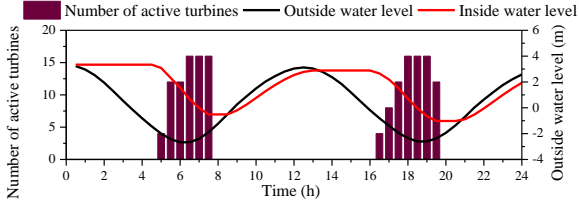

(b) Number of active turbines and water levels

Fig. 9. Operation scheme under price profile #1

TABLE I. COMPARISON OF ENERGY PRODUCTION AND REVENUE

| Operation scheme | Energy Production (MWh) | Revenue (£) |
|---|---|---|
| Max Electricity production (under price profile #1) | 519 | 153,485 |
| Max Revenue (under price profile #1) | 510 | 154,479 |

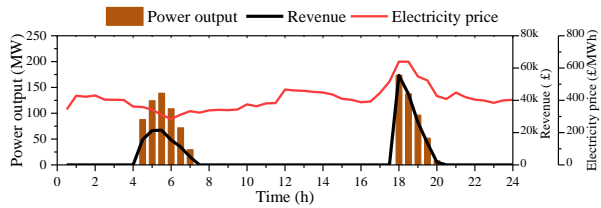

(a) Power output, revenue and price profile

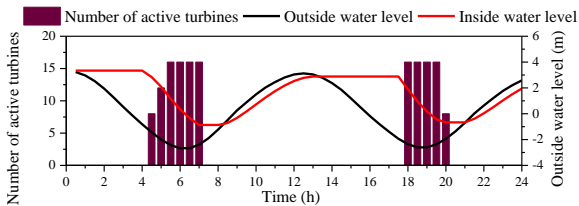

(b) Number of active turbines and water levels

Fig. 10. Operation scheme under price profile #2

### C. Quantification of energy storage capability of a tidal lagoon

When the water level inside the lagoon is higher than the outside, the tidal lagoon holds a certain amount of potential energy that could be converted into electrical energy. As long as the tidal lagoon maintains the high water level, its capability of generating remains and could be regarded as some sort of energy storage system. However, different from the battery, the level of energy stored in a tidal lagoon varies in different time periods as it depends on the seawater level outside the lagoon.

To explore maximum energy stored by the tidal lagoon in different time periods, we defined the level of storage in a tidal lagoon as the maximum electrical energy it could produce in one time step. The inner water level was set to the highest during all time periods while the sea level goes up and down, and the maximum possible generation output was calculated based on the varying head difference.

As shown in Fig. 11, the head difference has the same trend as in other figures, but with greater values. The level of storage is mostly proportional to the head difference. The maximum generation capacity of turbines limits the maximum power output at each time step. And whether the head difference is higher than the startup threshold determines if the tidal lagoon generates electricity or not. During most of the time, the tidal lagoon stores a certain amount of energy which can be provided to the power system. Although the storage capability of one single tidal lagoon is greatly impacted by the tides, the coordination of multiple tidal lagoons in different locations (operating under different tidal ranges) could help fill the valleys in energy storage, and provide the power system a continuous and reliable source of flexibility.

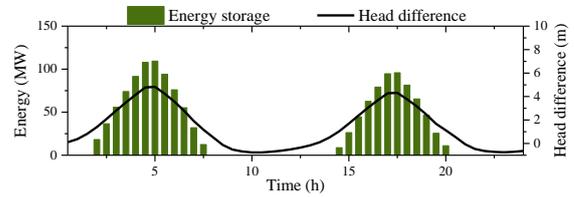

Fig. 11. The fluctuation of stored energy and head difference

### V. CONCLUSION

In this paper, an optimal operation scheme of a tidal lagoon was proposed to maximise its revenue from selling generated electricity. First the optimisation problem was modelled based on a 0-D tidal lagoon model and fully linearised to an MILP form for less computation complexity. Then the tidal lagoon's reaction to the fluctuating electricity prices was studied to provide more insight into how tidal lagoon could operate in an electricity market. Results shows that the tidal lagoon can adjust its power generation by controlling the number of active turbines in response to the energy prices, so more energy could be provided when it's most needed by the power system. The energy stored by the tidal lagoon was also evaluated and data show that the level of storage goes up and down with the ebb/flood tides outside the lagoon.

For future work, the study on tidal lagoon's operation scheme will be extended to the two-way generation and simulations will be conducted over a longer time period considering different tidal ranges on different days. Furthermore, the coordination of tidal lagoon fleet and other renewable energy sources will also be studied to see how tidal lagoons provide flexibility to the power system and gain revenues in the energy market.